# Thinned GaInP/GaInAs/Ge solar cells grown with reduced cracking on Ge|Si virtual substrates

Ivan García[1*], Laura Barrutia[1], Shabnam Dadgostar[2], Manuel Hinojosa[1], Andrew Johnson[3] and Ignacio Rey-Stolle[1]

[1]Instituto de Energía Solar, Universidad Politécnica de Madrid, 28040 Madrid, Spain
[2]GdS Optronlab, Universidad de Valladolid, 47011 Valladolid, Spain
[3]IQE plc, Pascal Cl, St. Mellons, Cardiff, CF3 0LW, United Kingdom

**Abstract**

Reducing the formation of cracks during growth of GaInP/GaInAs/Ge 3-junction solar cells on Ge|Si virtual substrates has been attempted by thinning the structure, namely the Ge bottom cell and the GaInAs middle cell. The theoretical analysis performed using realistic device parameters indicates that the GaInAs middle cell can be drastically thinned to 1000 nm while increasing its In content to 8% with an efficiency loss in the 3-junction cell below 3%. The experimental results show that the formation of macroscopic cracks is prevented in thinned GaInAs/Ge 2-junction and GaInP/GaInAs/Ge 3-junction cells. These prototype crack-free multijunction cells demonstrate the concept and were used to rule out any possible component integration issue. The performance metrics are limited by the high threading dislocation density over $2\cdot10^7 cm^{-2}$ in the virtual substrates used, but an almost current matched, crack-free, thinned 3-junction solar cell is demonstrated, and the pathway towards solar cells with higher voltages identified.

**1. Introduction**

The development of high efficiency III-V multijunction solar cells monolithically grown on silicon substrates is challenged by the large lattice mismatch and dissimilar thermal expansion coefficients between the III-V and Si materials, which causes the formation of cracks in the epilayers. Using compositionally graded buffers (CGB) has been demonstrated to be a way to achieve reduced threading dislocation densities (TDD) with GaAsP [1]–[4] and SiGe [5]–[8] – based structures, mainly. However, the thickness added to the structure by the CGBs makes it difficult not to surpass the critical thickness for crack propagation, experimentally observed to be around 5 μm for III-V on Si at typical growth temperatures [9]. Crack-free GaInP/GaAs/Si solar cells grown monolithically using GaAsP CGBs have been reported [10]. Due to the slightly higher than optimal bandgap of Si, this approach requires thinning the upper subcells to achieve current matching and maximizing the efficiency, which, at the same time, makes possible to prevent cracking.

In a parallel approach under development, TDD of $1\text{-}5\cdot10^6 cm^{-2}$ have been achieved with direct growth of Ge on Si [11]–[13]. This approach has the advantage of allowing the direct integration of already developed high efficiency solar cell structures based on Ge or GaAs substrates. In fact, promising results have been demonstrated for GaAs and GaInP single-junction solar cells [14]–[16]. For higher efficiency, thicker structures and more junctions are needed, and thus various approaches have been tested to reduce the formation of cracks. Scaccabarozzi et al. used coalesced Ge layers on deeply patterned Si substrates, with the surface morphology improved through chemical mechanical polishing (CMP). With this method, they attained GaInP/GaInAs/Ge|Si triple-junction solar cell (3JSC) exhibiting an external quantum efficiency (EQE) comparable to the samples grown on standard Ge substrates, but with a significant voltage loss attributed to an imperfect growth on the coalesced Ge template. Bioud et al. used embedded nanovoids at the Ge|Si interface by electrochemical porosification plus annealing, to achieve lower TDD, but which could be used also for reduced cracking [17]. Another approach by Oh et al. consisted on inducing the geometrically controlled formation of cracks by a notch pattern applied to the Si substrate, strategically arranged to enable the fabrication of the solar cell devices in the resulting crack-free regions [18]. All these methods intend to control the formation of cracks in structures over the critical thickness, so that their effect on the active regions of the solar cells is limited. This has the advantage of decoupling the strategy to mitigate the effect of the cracks from the design of the solar cells. On the other hand, the modifications to the substrate required, often involving nanopatterning, add complexity to the fabrication process, which could hinder the application to mass production scenarios.

Conversely, the approach that we investigate in this work pursues limiting the formation of cracks by redesigning the solar cell structure, in the context of GaInP/GaInAs/Ge 3JSC. In our previous works, we showed that standard 3JSC structures (over 6 micron thick) could be grown on Ge|Si templates with a 5 µm thick Ge layer, but the formation of cracks contributed to the low performance of the solar cells fabricated [19], [20]. In this work we analyze the thinning of the 3JSC structure to reduce the formation of cracks. Firstly, we use thinner Ge layers of 3 µm in the Ge|Si template, which are enough to produce the required photocurrent for a current-matched 3JSC [19]. Then, we test the effect of aggressively reducing the thickness of the GaInAs middle cell. A side benefit is that the lower recombination volume allows to obtain higher voltages in these thinned solar cells grown on Ge|Si templates. However, this thinning causes a reduction in the photocurrent of the middle cell, losing the current-matching condition in the 3JSC. Adding a distributed Bragg reflector (DBR) below the GaInAs middle cell, similarly as in space solar cell applications to increase the radiation hardness of the cells [21], would only allow thinning this subcell to about 2500 nm without photocurrent losses, and the added thickness by the DBR would defeat the objective of achieving a globally thinner structure. A different approach is followed in this work: the reduction of the middle subcell bandgap to increase its photocurrent is explored theoretically and demonstrated experimentally to allow regaining current matching in the 3JSC. The voltage reduction due to the lower bandgap in the GaInAs subcell represents a significantly lower loss for the 3JSC performance than the effect of the current mismatch when this subcell is only thinned without changing its bandgap. Nevertheless, this voltage drop represents a performance loss in comparison to the standard 3JSC. The tradeoff between the advantages of using Ge|Si substrates (lower usage of Ge, potential lower cost, possibility of substrate detachment using embedded porous layers [19], [20], etc) and the lower efficiency will determine the economic viability of this approach.

In this paper we first present a theoretical assessment of the effects of thinning the structure on the performance of a 3JSC, and the potential of using a lower bandgap in the middle cell to mitigate the associated losses. Then we present the experimental results obtained with thinned GaInAs middle cells grown on Ge|Si virtual substrates, compared to those grown on standard Ge substrates. By increasing the In content in this subcell we show that the photocurrent of the thinned subcell can be recovered, at the expense of some voltage loss. Finally, we implement a proof-of-concept 3JSC, including the thinned GaInAs subcell, which shows no significant cracking and no component integration issues.

### 2. Methods

The fit to the experimental EQE and calculation of the short circuit current densities ($J_{sc}$) is performed using the Scattering Matrix Method [22] and Hovel models, with initial minority carrier parameters measured previously in our structures [23]. The calculation of the open circuit voltages ($V_{oc}$) for the GaInAs subcells with different In contents, discussed in Section 3, was performed following the model by Steiner et al. in [24]. The inputs to this model are the optical properties of the semiconductor stack, the EQE and the internal luminescence efficiency ($\eta_{int}$), which defines the ratio of global radiative to total recombination in the absorber (emitter + base) and represents the electronic quality of the material. Values of ~0.4, observed in our standard GaInAs solar cells, are used. This $\eta_{int}$ value is higher than expectable for the so far lower quality cells grown on Ge|Si substrates, but the theoretical study presented here intends to analyze the potential performance variations of the cells when redesigned by changing their thickness and bandgap, assuming at least the same minority carrier properties as in cells grown on standard Ge substrates. Note that this model captures the decrease in $V_{oc}$ as the bandgap decreases and also as the absorber thickness increases, due to a higher recombination in the volume [24]. As will be discussed later, this effect will play a significant role in determining the voltage loss as the solar cell bandgap and thickness are redesigned. The $J_{sc}$ calculation from EQE data is performed using the AM1.5G spectrum (1000 W/m$^2$). The trends obtained for the direct and AM0 spectrum are analogous.

The Ge|Si virtual substrates were fabricated at IQE PLC using silicon substrates with a miscut of 6º towards the nearest (111) plane. The Ge layers were deposited by CVD as explained elsewhere [19]. In this work, virtual substrates with a 3 μm thick Ge layer are used, as compared to our previous reports for 5 μm thick Ge layers [19], [20]. These thinner Ge subcells were experimentally confirmed to achieve the required $J_{sc}$ for the 3JSC application. The virtual substrates used exhibit a threading dislocation density (TDD) of ~2·10$^7$ cm$^{-2}$, measured by plan-view cathodoluminescence mapping (CL) of GaInAs overbuffer layers grown on them. For comparison, standard Ge substrates were also used. They were p-type doped with Gallium and again with a miscut of 6º towards the nearest (111) plane. The III-V epistructures were grown in a research-scale, horizontal, low-pressure AIX200/4 reactor, using AsH$_3$, PH$_3$, TMGa, TMIn, TMAl, DMZn, DTBSi and DETe as precursor molecules. The growth rates used vary from 2 to 4 μm/hr and the growth temperature is 675ºC for phosphides, 640ºC for arsenides and 550ºC for tunnel junctions. The 3JSC structure used is depicted in Figure 1. Further details of these baseline structures can be found elsewhere [25]. In this work we use the same growth conditions for the structures grown on Ge|Si as in our standard structures on Ge substrates. Reducing the growth temperature has the potential of reduce cracking, and it is planned to be studied in future works.

The lattice-constants and compositions in the structures grown were characterized by High Resolution X-Ray Diffraction (HRXRD) using an X'Pert Panalytical MRD diffractometer with the x-ray beam conditioned using a 4xGe(220) monochromator, to obtain rocking curves and reciprocal space maps (RSM). Cathodoluminescence (CL) measurements were taken using a LEO 1530 (Carl-Zeiss) field- emission scanning electron microscope (FESEM) equipped with a MonoCL2 (Gatan UK) CL system. Panchromatic images were recorded with a photomultiplier tube (PMT) for the visible range, and with an InGaAs detector for the near infrared. The images were acquired for different e-beam energies, from 5 keV to 25 keV, to reveal the defects of the different layers forming the structure. All the measurements were done at 80K to enhance the CL emission intensity.

Solar cell manufacturing is based on standard photolithography techniques for the definition of the front grid, which is based on the AuGe/Ni/Au system. An inverted square geometry is used for the front grid, with an active area of 0.1 cm$^2$. The back contact is based on gold for solar cells fabricated on Ge substrates, and Pd/Ti/Pd/Al for cells fabricated on Si/Ge virtual substrates, which allows obtaining high performance contacts to the Si substrate without applying detrimental thermal loads to the III-V structure [26].

EQE and reflectivity (R) were measured using a custom-built system based on the lock-in technique to detect the photocurrent of the solar cell excited by monochromatic light obtained using a Xe lamp and a grating monochromator. A monitor cell is used to compensate for fluctuations in the Xe-lamp. I-V curves were obtained using a Xe-lamp solar simulator and a 4-quadrant Keithley source-monitor unit. Finally, bandgap variations were measured by spectral electroluminescence (EL) in the 300-1100 nm range using a calibrated fiber spectrometer based on a silicon CCD detector. For the Ge subcells, the bandgap is assumed to be 0.65 eV .

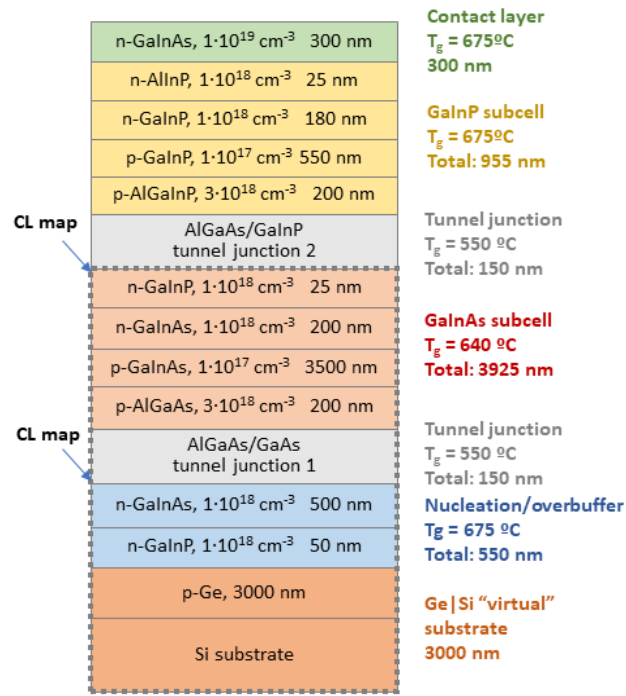

Figure 1. Structure of the baseline 3JSC grown on a Ge|Si virtual substrate. The dashed line frame indicates the GaInAs/Ge double-junction solar cell (2JSC) structure that is redesigned along this work. The arrows indicate where CL maps were taken to quantify the TDD mentioned along this paper.

**3. Theoretical analysis.**

The proposed approach based on thinning and increasing the indium content of the GaInAs subcell allows to substantially reduce the thickness of the 3JSC structure but at the same time reduces the $V_{oc}$ of the GaInAs subcell, due to its lower bandgap. Besides, the lower bandgap in this subcell implies less spectral range available to the Ge bottom cell. We quantify these two effects with this theoretical study. A GaInP top cell grown lattice-matched to the metamorphic GaInAs subcell with higher indium content as in upright metamorphic 3JSC designs [27], would have a lower bandgap than GaInP lattice-matched to Ge, which implies an additional $V_{oc}$ loss in the 3JSC. This can be mitigated by using AlGaInP instead, in view of the excellent photovoltaic quality obtained by other authors in single [28] and multijunction solar cells [29]. Therefore, in our theoretical study, the bandgap of the top cell is assumed to have its usual value for standard lattice-matched 3JSC designs.

The $J_{sc}$ of the GaInAs and Ge subcells in a 3JSC, and the $V_{oc}$ of the GaInAs subcell were computed for a range of GaInAs subcell base layer thicknesses and In contents. The effect of the In content on the absorption coefficient is estimated using an square root model applied to the fitted GaAs data [30]. The results shown in Figure 2-A illustrate the change in the EQE of the middle and bottom subcells as the GaInAs base layer thickness and bandgap are decreased. Graphs B to D are contour plots of parameters vs the GaInAs subcell thickness and In composition. The GaInAs subcell $J_{sc}$ shown in graph B clarifies the amount of In to add to the GaInAs material to keep the $J_{sc}$ constant as its thickness is decreased. The noticeable transition from a slower to a higher increase rate of the In content required as the thickness is decreased takes place when the GaInAs subcell bandgap and EQE low energy edge enters one of the characteristic dips in the solar spectrum (see graph A). In fact, this inflexion point is not present for the AM0 spectrum (not shown), for which lower In contents are required to maintain the Jsc in thinned GaInAs subcells. Around 8% (~0.5% lattice mismatch to the Ge substrate) In content is required for a thickness of 1000 nm (compared to ~1% In for the optically thick 3500 nm baseline). This low In content should not be a problem to produce high photovoltaic quality material. The method used to implement this GaInAs composition is presented in Section 4.2.

Figure 2-C shows the evolution of the $J_{sc}$ in the Ge subcell, revealing the expected opposite behaviour as compared to the GaInAs subcell: the thinner and lower In content in the GaInAs, the more photons are transmitted to the Ge junction and the higher is its $J_{sc}$. Besides, the shape of the contours expose and interesting effect: the $J_{sc}$ of the Ge subcell decreases even if the $J_{sc}$ of the GaInAs subcell is kept constant as the thickness and In content change simultaneously, as occurs by following the iso-$J_{sc}$ dashed line. In fact, the standard 3JSC structure used as baseline in this study is designed with an optically thick GaInAs subcell, and with other GaInAs layers underneath, including the tunnel junction and overbuffer (see Figure 1). As the GaInAs subcell is made thinner, these other GaInAs layers underneath start to parasitically absorb photons, explaining the $J_{sc}$ decrease in the Ge bottom cell. In any case, the $J_{sc}$ in the Ge subcell for the 1000 nm and 8% In content GaInAs subcell is over 18 mA/cm$^2$, which is higher than required for a current-matched 3JSC (~15.1 mA/cm$^2$), although the effect of dislocations in Ge subcells fabricated with Ge|Si virtual substrates is expected to reduce this $J_{sc}$ excess [19].

Finally, the calculated $V_{oc}$ in the GaInAs subcell for higher In content and thickness can be observed in Figure 2-D. Some relevant $V_{oc}$ variations from this contour plot are summarized in Table 1. The $V_{oc}$ drop when increasing the In concentration from 1% to 8% (for constant thickness), is estimated to be 94 mV. If the thickness is reduced as needed to keep the $J_{sc}$ constant, the $V_{oc}$ drop decreases to 70 mV, i.e., the $V_{oc}$ recovers partially as the base layer is made thinner. It is important to point out that, as detailed in the Methods section, this study assumes a $\eta_{int}$ of 0.4, which corresponds to our standard material quality but it is low as compared to state-of-the art GaAs [31]. With a higher $\eta_{int}$, the effect of thickness would be lower. However, despite the best TDD achieved in Ge|Si virtual substrates is in the 1-5·10$^6$cm$^{-2}$ range and with expectations to obtain even lower values, some effect of the TD on the $\eta_{int}$ is likely. Previous analysis on high quality metamorphic GaInAs with TDD values in this range showed $\eta_{int}$ of 0.3-0.6 at 1-sun current levels [32]. Therefore, we can consider the $V_{oc}$ drop estimation obtained here as good estimation for GaInAs cells grown on Ge|Si virtual substrates with the TDD range considered. However, note that, as will be shown in next sections, the effect of thickness on the $V_{oc}$ of the experimental GaInAs subcells studied here largely exceeds these estimations due to the higher TDDs in the Ge|Si virtual substrates used in this work.

In summary, thinning and increasing the In content in the GaInAs subcell will produce an estimated $V_{oc}$ loss of 70 mV in the 3JSC. This represents a relative ~3% lower efficiency at 1 sun and relative 2% for high concentration (1000 suns and assuming an ideality factor of 1 in all junctions to extrapolate the $V_{oc}$ to this concentration). The effect of the TDD, inherent to the use of Ge|Si virtual substrates, can produce additional $V_{oc}$ losses. For the 1·10$^6$ cm$^{-2}$ TDD range, $V_{oc}$ drops of 70 mV have been reported in GaInP/GaAs 2JSC with usual thicknesses, being the GaAs subcell responsible for ~60 mV [33]. Note however that our thinned GaInAs subcell would be less affected by the TD and, therefore, the $V_{oc}$ drop would be lower. No estimations of the effect of the TD on the Ge subcell, which is also affected by misfit dislocations, are available, but the experimental results on highly dislocated Ge|Si virtual substrates, compared to standard Ge

substrates (shown in Section 4.1, Table 3), indicates that this subcell can contribute with a few tens of mV more to the losses. Therefore, the effect of TDD in the $1 \cdot 10^6$ cm$^{-2}$ range can account for a significant increase in the $V_{oc}$ loss but they can be minimized as long as better quality Ge|Si substrates are achieved, conversely to the impact of redesigning the structure on the $V_{oc}$ that is inevitable. Eventually, the aggregated economic benefits of using high quality Ge|Si substrates must be higher than the losses produced by a 2-3% lower efficiency, to make this approach economically viable, which seems feasible.

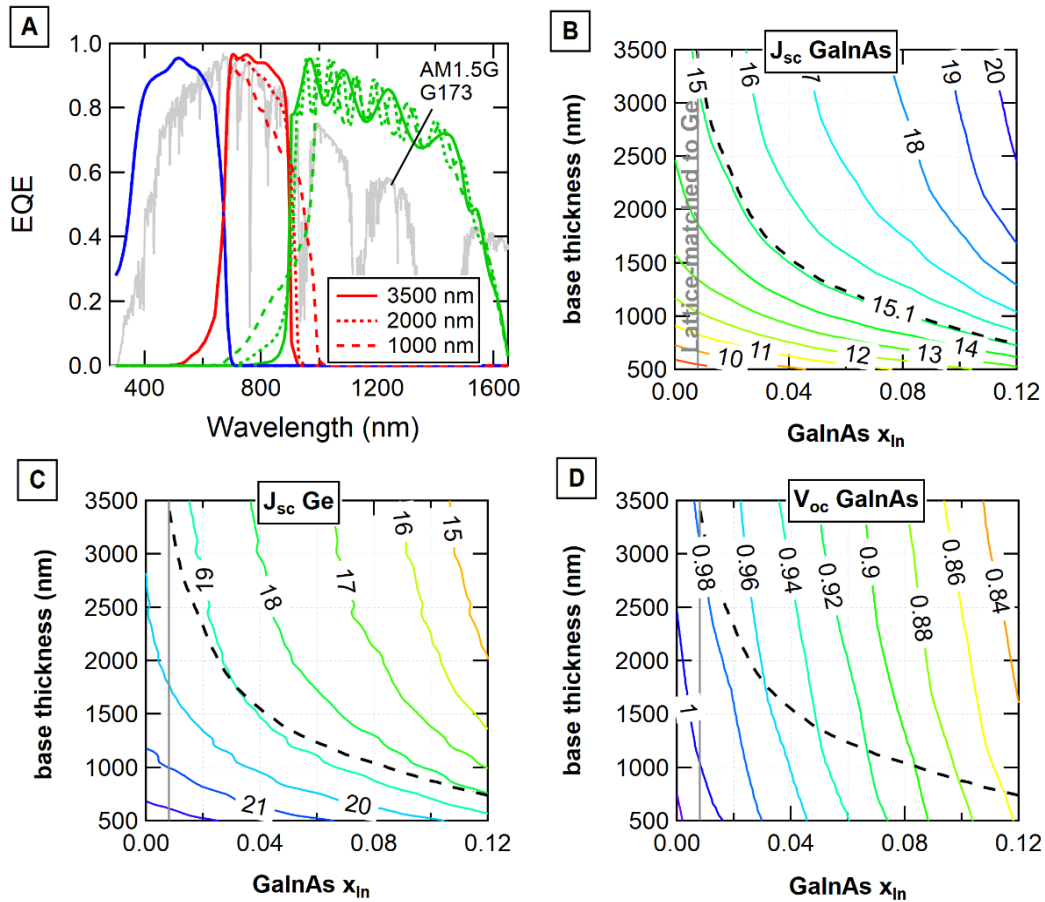

Figure 2. A: modeled EQE of the 3JSC subcells, for three GaInAs base thicknesses and In compositions that produce current matching with the top GaInP subcell. The grey curve is the AM1.5G solar spectrum in arbitrary units. B to D: $J_{sc}$ of the GaInAs subcell, $J_{sc}$ of the Ge subcell and $V_{oc}$ of the GaInAs subcell vs its base thickness and In composition. The dashed black lines in graphs C and D are the iso-contour of GaInAs subcell $J_{sc}$ at the value obtained for the standard case. The solar cells have a MgF$_2$/ZnS (85/40 nm) anti-reflection coating that optimizes the $J_{sc}$. The AM1.5G (1000 W/m$^2$) solar spectrum is used in all cases.

| GaInAs % In | GaInAs base thickness (nm) | ΔV$_{oc}$ (mV) |
|---|---|---|
| 1.00 | 3500 | 0 |
|  | 2000 | +12 |
|  | 1000 | +23 |
| 8.00 | 3500 | -94 |
|  | 2000 | -82 |

| | 1000 | -70 |
|---|---|---|

Table 1. Calculated GaInAs cell $V_{oc}$ variation with respect to the standard case (1% In, 3500 nm), for the In contents and thicknesses explored. The target GaInAs subcell design is highlighted in bold.

## 4. GaInAs subcell experimental results.

As commented before, the GaInAs subcell is the thickest component of the 3JSC structure and, therefore, a target in our efforts to reduce the total thickness and prevent cracking. To this end, GaInAs/Ge double-junction solar cells (2JSC) were grown, processed and characterized to examine the effect of thinning the base layer and changing its bandgap by increasing the In content.

### 4.1. Effect of thinning the base layer.

We first focus on the effect of thinning the base layer of the GaInAs subcells grown on Ge|Si virtual substrates. Using the baseline structure shown in Figure 1, GaInAs/Ge 2JSC were grown with different base layer thicknesses while keeping the rest of the structure nominally identical, and solar cells were fabricated with these structures. The EQE of these solar cells is shown in Figure 3, which also includes the case of a 3500 and 1000 nm base layer thickness for GaInAs cells grown on a Ge substrate, as a reference. The EQEs of the cells grown on a Ge substrate were fitted and the obtained parameters were used for the initial fitting of the cells grown on Ge|Si. A summary of the extracted diffusion lengths is shown in Table 2, as well as the $J_{sc}$ calculated using these EQE with 700 nm as high energy cut-off, emulating the presence of a GaInP top cell as in a 3JSC. Figure 3 also includes two representative Nomarski microscope images of structures grown on Ge|Si corresponding to 3500 and 1000 nm base layer thicknesses, which show a high density of cracks in the thick structure, and no cracks present in the thin case

First, a significant reduction in the carrier collection efficiency, and modeled diffusion lengths, can be observed when comparing the solar cells with standard thickness grown on Ge and Ge|Si substrates. The high TDD in the Ge|Si templates used can justify this. On the other hand, the higher diffusion length observed as the base layer is thinned in the GaInAs subcells grown on Ge|Si substrates was unexpected. To further investigate this finding, CL maps were taken on the samples with 1000 and 3500 nm base thicknesses for two cases: first, structures with only the top contact layer etched, and second, with all the layers etched down to the GaInAs overbuffer (measurement positions indicated with arrows in Figure 1). The resulting CL images for the 3500 nm thick base cell are shown in Figure 3. Firstly, the CL map for the GaInAs overbuffer shows the TD in a density of $\sim 2 \cdot 10^7$ cm$^{-2}$, with no other remarkable feature. However, the CL map taken at the top of the GaInAs cell shows an unexpected array of misfit dislocations. All the layers in these structures are lattice-matched to Ge, as corroborated by HRXRD RSM. It seems that the TD coming from the substrate, or new TD, glide during the growth of the GaInAs subcell, generating misfit dislocations (MD). The driving force for this dislocation movement is not certain, but thermal strain, due to the thermal mismatch between Si and the rest of the structure, is the expected cause. The Ge|Si template fabrication includes thermal cycling (at temperatures higher than used for the growth of the III-V structure) for the minimization of the TDD. However, as thickness is added to the structure during the growth of the solar cells, additional thermal strain can be originated, and drive the movement and/or nucleation of existing TD. The CL maps for the 1000 nm base layer cell showed a similar TDD in the overbuffer layer and misfit array density in the GaInAs (not shown in Figure 3). For the CL beam energy used, the scan integrates a depth of 700 nm. Therefore, it appears that the formation of misfits starts for as thin GaInAs layers as < 1000 nm. Concerning the possible role of cracks, as the base thickness is reduced, the crack density (summarized in Table 2) decreases, with only a few cracks at the very edges of the 4" wafers observed for the thinnest samples (achieving the primary objective of this study). The effect of the macroscopic cracks on the carrier collection should be limited for the crack densities measured but it is reasonable to expect the presence of related defects (microcracks for example), which could have an impact on the EQE, but are removed as the structure is thinned. This can explain the higher carrier collection efficiencies for thinner GaInAs samples. At any rate, beyond the elimination of cracks that is demonstrated for the thinned samples, it seems obvious that the unexpected formation of MD in the active layers must be prevented. We expect that using Ge|Si substrates with a lower TDD will be decisive in this respect.

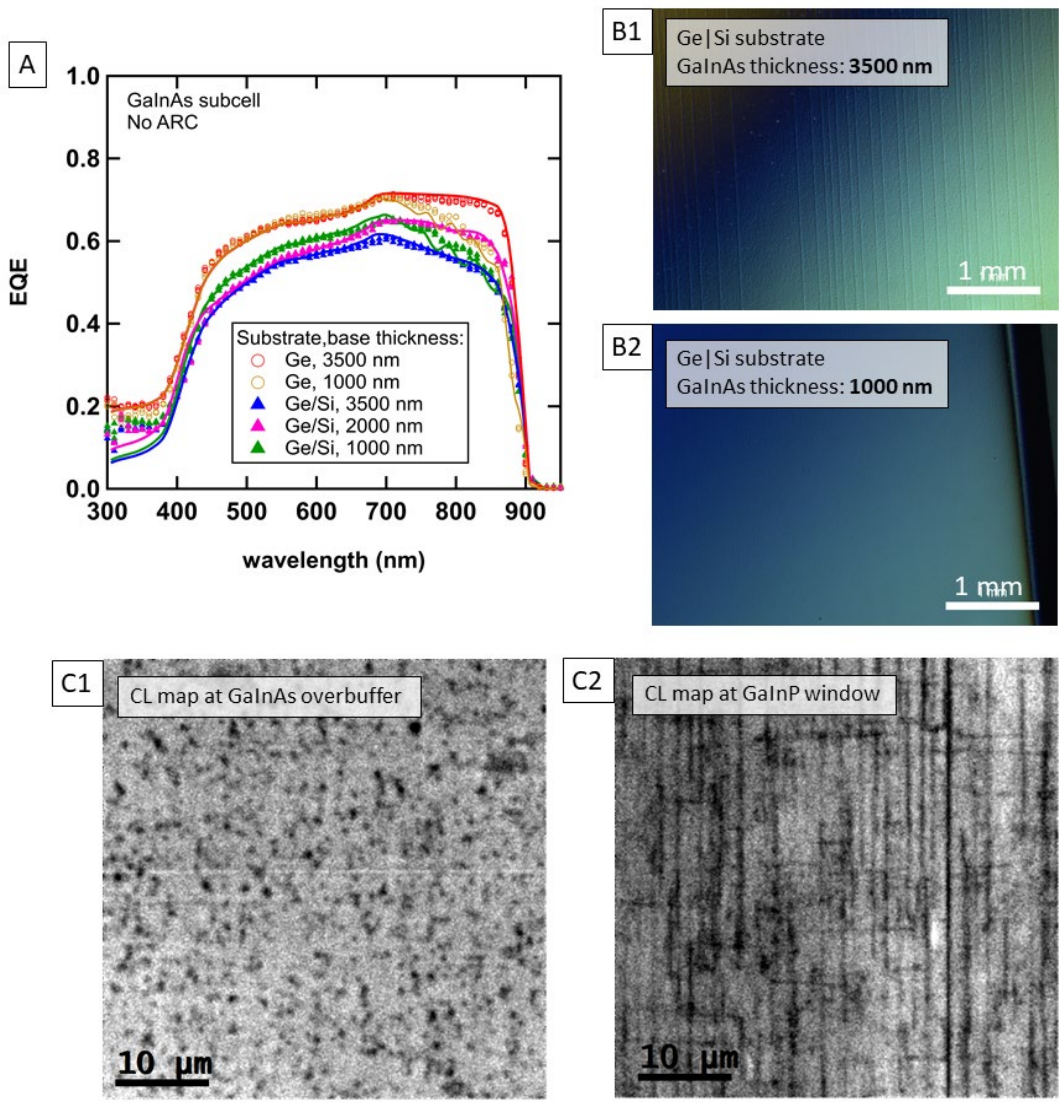

Figure 3. A: EQE of lattice-matched (1% In) GaInAs subcells grown on Ge and Ge|Si virtual substrates with different base thickness. Symbols: experimental; lines: fit using Hovel model. B1-2: Nomarski images of the sample surface for the standard (B1) and thinned (B2) structures grown on Ge|Si substrates. C1-2: CL map images (taken at 10 keV) of the cell structure with a 3500 nm base layer taken at the GaInAs overbuffer (C1) and GaInP window layer of the GaInAs middle cell (C2).

| Substrate | Base thickness (nm) | % In | $L_{h,emitter}$ (μm) | $L_{e,base}$ (μm) | $J_{sc}$ (>700nm) (mA/cm$^2$) | average cracks/mm |
|---|---|---|---|---|---|---|
| Ge | 3500 | 1 | 2.0 | 20.0 | 8.7 | 0 |
|    | 1000 | 1 | 2.0 | 20.0 | 7.8 | 0 |
| Ge\|Si | 3500 | 1 | 0.3 | 1.8 | 6.9 | > 5 |
|        | 2000 | 1 | 0.3 | 5.1 | 7.9 | 0-5 |
|        | 1000 | 1 | 0.4 | 16.1 | 7.4 | ~0 |
|        | 1000 | 6.7 | 0.4 | 2.0 | 8.5 | ~0 |

Table 2. Summary of minority carrier diffusion lengths, $J_{sc}$ and crack density of the samples presented along this work. $L_{h,emitter/base}$ : diffusion lengths of minority carriers in the emitter (holes) and base (electrons) obtained by fitting the EQE curves in Figure 3 (for GaInAs with 1 % In) and Figure 6 (for GaInAs with 6.7% In). The $J_{sc}$ was calculated using the AM1.5G spectrum and the EQE for a high energy cut-off of 700 nm. The rightmost column shows the average linear density of cracks measured.

Light J-V curves of the GaInAs/Ge 2JSC were measured on a statistically significant number of devices and are shown in Figure 4. The multiple lines with same color in Figure 4 give a visual idea of device-to-device variability. The dispersion observed is partly caused by a different metal shadowing in the devices (varied number of fingers in the front contact, including devices with no fingers for the EQE measurements). The $V_{oc}$ and $W_{oc}$ (difference between $V_{oc}$ and bandgap) obtained are summarized in Table 3. The $V_{oc}$ of the Ge subcells shown in Table 3 was measured by first etching the top GaInAs junction and then processing the Ge cells as single-junction devices. The relatively low $V_{oc}$ measured in the Ge subcell fabricated with standard Ge substrates is caused by the thermal load applied during the growth of the rest of the structure, which degrades the emitter and front surface minority carrier properties of the Ge junction [34]–[36]. However, the difference in thermal load between the 2JSC with 3500 and 1000 nm base thickness in the GaInAs subcell was observed to cause variations in the $V_{oc}$ of less than 5 mV, and therefore are neglected in this study for simplicity.

As expected, the $J_{sc}$ in the cells grown on standard Ge substrates decreases significantly when decreasing the absorber thickness from 3500 to 1000 nm, in agreement with the EQE (Figure 3). A noticeable increase in $V_{oc}$ of 26 mV is observed, too, which agrees well with the theoretical results shown in Table 1. The $J_{sc}$ trends for the GaInAs/Ge cells grown on Ge|Si also match the expected values from the EQE measurements, improving as the absorber is thinned to 2000 nm thanks to a better carrier collection, but decreasing again for 1000 nm as the reduction in optical absorption dominates. The $V_{oc}$ increases for thinner absorbers, but with quantitative variations (up to almost 100 mV) exceeding the theoretical predictions (~23 mV) for standard Ge substrates, given the much worse material quality obtained on the Ge|Si templates. In addition, the achievement of a better material quality in thinner layers, as observed in the EQE responses and discussed before in this section, might contribute to this effect too. The $V_{oc}$ difference between the thin GaInAs subcells grown on Ge|Si and on Ge (218 mV) agrees with previously reported results using GaAsP/Si templates with similar TDD [10].

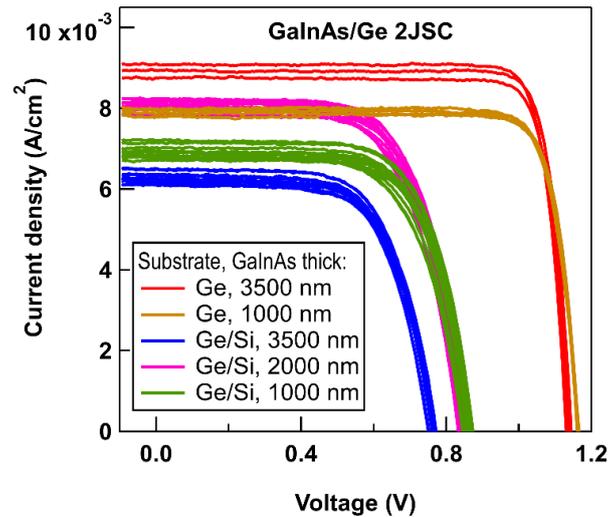

Figure 4. Light J-V curves of the lattice-matched GaInAs (1% In)/Ge 2JSC grown using standard Ge and virtual Ge|Si substrates, for different thicknesses of the GaInAs absorber layer. The irradiance was set by using a GaAs reference cell and filtering the photons energies above 700 nm, as done for the calculation of the $J_{sc}$ from the EQE listed in Table 1.

| Substrate | Subcell | Base thickness (µm) | % In | Subcell $V_{oc}$ (V) | Subcell $W_{oc}$ (V) |
|---|---|---|---|---|---|
| Ge | Ge | 175 | - | 0.220 | 0.450 |
| Ge\|Si | | 3 | - | 0.140 | 0.530 |
| Ge | GaInAs | 3.5 | 1 | 0.918 | 0.472 |
| | | 1.0 | 1 | 0.944 | 0.446 |
| Ge\|Si | | 3.5 | 1 | 0.622 | 0.768 |
| | | 2.0 | 1 | 0.702 | 0.688 |
| | | 1.0 | 1 | 0.726 | 0.664 |
| | | 1.0 | 6.7 | 0.560 | 0.740 |

Table 3. Experimental $V_{oc}$ of the Ge and GaInAs subcells presented along this work, obtained from the J-V curves shown in Figure 4 (for GaInAs with 1 % In) and Figure 7 (for GaInAs with 6.7 % In). For the GaInAs subcells, it is obtained by using the $V_{oc}$ of the GaInAs/Ge 2JSC and subtracting the voltage of the Ge subcell in each case. The $V_{oc}$ values used are the average for the samples measured. The $W_{oc}$ is the difference between the bandgap and the $V_{oc}$ of the subcell.

Therefore, we have shown that drastically thinning the base layer of GaInAs subcell grown on Ge|Si substrates down to 1000 nm eliminates the formation of macroscopic cracks and recovers its performance by a significant factor. The J-V curve performance metrics of the GaInAs subcell are still poor, given the high defect density of the Ge|Si templates used, and the unexpected formation of misfit dislocations. Improving the GaInAs material quality by using lower TDD substrates is a priority, but in the event that the high quality material grown on standard Ge substrates could not be attained, thinning the subcells will aid not only in reducing cracking but also in recovering the $V_{oc}$.

**4.2. Effect of reducing the bandgap.**

Reducing the bandgap of the GaInAs subcell should allow to reduce its thickness while preserving the current in the 3JSC, at the cost of some voltage loss, as shown in Section 3. In order to reduce the GaInAs base layer to 1000 nm, the indium content has to be increased to ~8%, with a lattice mismatch to the substrate of ~0.5%. This lattice constant transition can be achieved using a compositionally graded buffer (CGB). However, the CGB adds thickness to the semiconductor structure, defeating the original objective of thinning the 3JSC solar cell. Therefore, we have used a CGB design with a minimum increase in the thickness by integrating it in the overbuffer layer grown on top of the GaInP nucleation layer (see Figure 1). This overbuffer, typically 500 nm of GaInAs, is used to separate the Ge substrate from the active III-V layers and prevent the effect of Ge solid state diffusion reaching active layers [36]. The first CGBs implemented had the same thickness but it was found difficult to attain a proper strain relaxation. To improve it, a 250 nm overshoot layer with a higher In content than the target was added to the structure, as shown in Figure 5. The TJ and BSF layers also play the role of step-back layer in usual CGB structures. A negligible residual strain in the absorber GaInAs layer was achieved, as corroborated by the HRXRD reciprocal space maps taken, shown in Figure 5. As can be seen, the CGB designed does not follow a linear grading, but uses a step jump at the beginning to enhance earlier relaxation [37]. With this CGB an In composition of 6.7% was achieved and was used to demonstrate the concept. Further fine tuning of this CGB adjusting the In content can be easily achieved, if required.

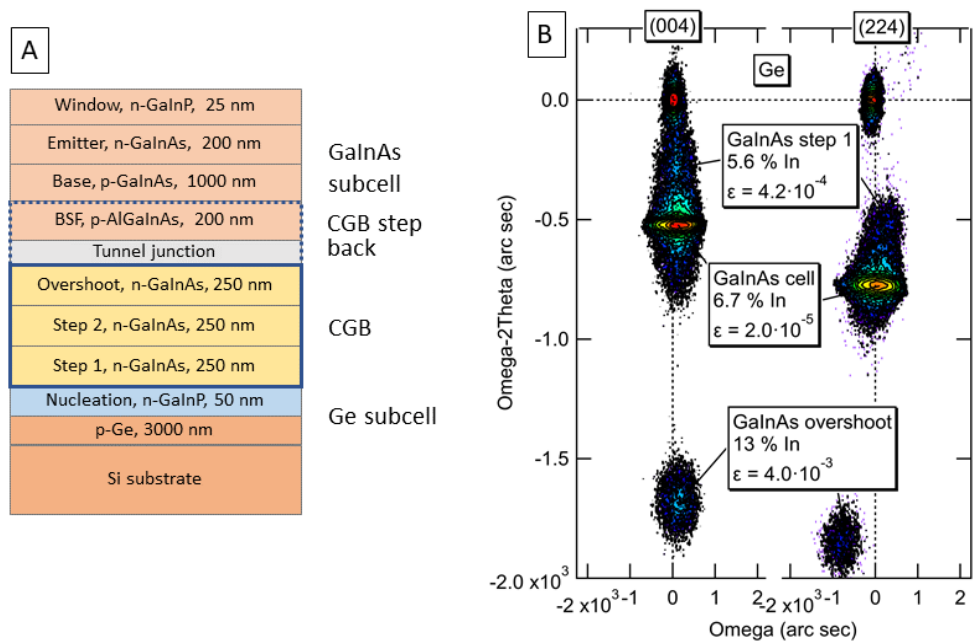

Figure 5. <u>A</u>: structure of a metamorphic GaInAs(6.7% In)/Ge 2JSC including a CGB that replaces the overbuffer layer. <u>B</u>: symmetric and asymmetric RSM taken to measure the lattice constants and derive the residual strain ($\varepsilon$) and composition of the layers in the CGB and GaInAs cell grown on Ge.

Figure 6 shows the surface of the CGB + GaInAs subcell structures under the Nomarski microscope. No visible cracks can be seen, and the expected crosshatch from the growth with relaxation of the CGB is present. The EQE of the cells with standard and increased In content are shown also in Figure 6. The cut-off energy increase (~88 meV) agrees with the change of the In content from 1% to 6.7% in the GaInAs absorber. The fit to the EQE for this case (solid line) was achieved using a lower electron diffusion length in the base layer than in the case of the same cell thickness but 1.0% In, as detailed in Table 2. This lower material quality is attributed to additional TD and roughness introduced during the relaxation of the CGB structure.

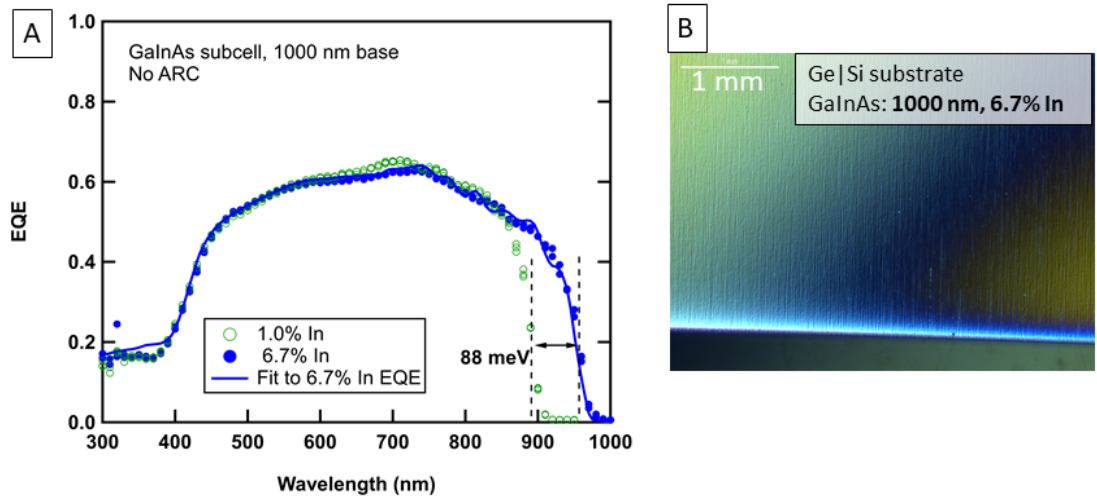

Figure 6. <u>B</u>: Nomarski microscope image of the as-grown surface of metamorphic GaInAs(6.7% In)/Ge 2JSC on Ge|Si. <u>A</u>: measured EQE of the GaInAs cells grown on Ge|Si with 1% and 6.7% In content, and 1000 nm thickness in both cases. The solid line is the fit to the EQE of the 6.7% In GaInAs cell.

Light J-V curves of these cells are shown in Figure 7, again for a statistically significant number of devices. The expected increase in $J_{sc}$ is observed for the cells with higher In content, but the $V_{oc}$ decrease (166 mV, see Table 2) is higher than corresponding to the lower bandgap (88 mV). Therefore, these J-V curves demonstrate functional thin and crack-free 2JSC, with a $J_{sc}$ approaching the values for thick GaInAs cells grown on standard Ge substrates. The electronic quality and $V_{oc}$ must be improved to obtain the targeted device performance. Using Ge|Si virtual substrates with reduced TDD will contribute largely to this objective.

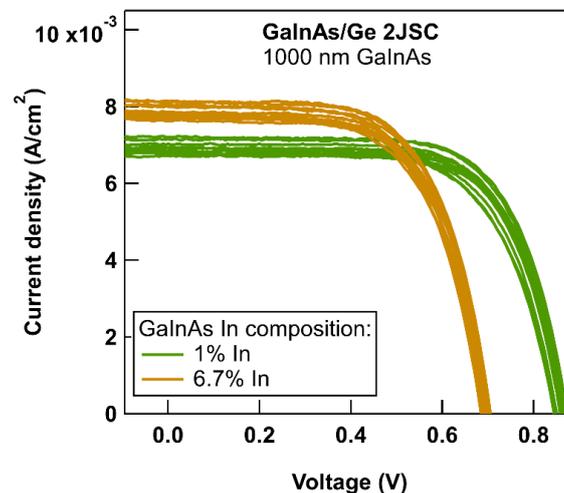

Figure 7. Light J-V curves of the GaInAs/Ge 2JSC cells grown on Ge|Si with 1% and 6.7% In content, and 1000 nm thickness GaInAs.

## 5. GaInP/GaInAs/Ge|Si triple-junction solar cell results.

The approach of thinning and decreasing the bandgap of the GaInAs subcell makes sense if current matching in the 3JSC can be preserved by either thinning the GaInP top cell or using AlGaInP to raise its bandgap. However, at this stage of development we have not attempted to optimize the top cell yet. The 3JSC implementations so far are not intended to achieve a high efficiency, but to demonstrate a crack-free structure, anticipate potential component integration issues and assess the performance obtained.

During the development iterations to build a functional 3JSC with thinned and higher In content GaInAs subcell, the compositions of the layers above this subcell had to be adjusted to be lattice-matched. This included the sensitive high bandgap tunnel junction, which uses a quantum-well to improve the tunneling probability [38] and resulted to work adequately for the new compositions. Besides, no significant cracking was observed in the as-grown 3JSC structures, as shown in Figure 8. In addition to the crossh-hatch from the metamorphic growth, macroscopic defects can be observed in these Nomarski images, which are caused by particles formed during dicing of the Ge|Si virtual substrates prior to growth. These deffects resulted in some devices being shunted, which were discarded. The fact that no cracking was observed in these 3JSC samples was surprising since, with a total thickness of ~6 μm, they are slightly above the expected critical thickness for the formation of cracks [9]. In fact, the GaInAs/Ge cells with such thicknesses showed cracks, as presented in previous sections. The absence of cracks in the 3JSC structures can be explained by the effect of the residual compressive strain in the CGB and GaInAs subcells, which can partially compensate the tensile strain during cooldown. This effect might provide advantage by intentionally modulating the residual compressive strain in the GaInAs subcell and CGBs, which we are investigating currently. After processing of the epistructures into solar cell devices, some cracks appeared, although still is a much lower density than in the thick GaInAs/Ge cells, as illustrated in Figure 8. This means that some residual strain is locked in the structure and is released during manipulation for the processing. We are currently investigating the effect of this residual strain on the reliability of the devices, and ways to minimize it by redesigning the CGB. Further thinning of the structure will, of course, be helpful for this purpose. Ge|Si substrates with thinner Ge layers will be tested in the near term.

Figure 8 shows the EQE of the top and middle cells in the 3JSC, compared to the case of a typical design on standard Ge substrates. The Ge bottom cell is not shown because it was not possible to measure it due to its leaky behaviour similarly as already reported for Ge|Si substrates with 5 μm Ge layer [19], [20]. As compared to the standard design, the GaInP top cell shows the expected shift in absorption edge in both the absorber material and window layer as a result of the higher In content. The EQE of the GaInAs subcell is similar to that shown for the GaInAs/Ge 2JSC in Figure 6 but filtered by the GaInP top subcell and upper tunnel junction.

The potential of this solar cell design to attain the required $J_{sc}$ once the top cell is appropriately designed was analyzed. Using the same minority carrier parameters as obtained from the fits to the experimental data, the EQE of the 3JSC was calculated for a top cell with the same bandgap as in the standard Ge substrate case (i.e., using AlGaInP) and optimized thickness, and with an anti-reflection coating (ARC) based on $MgF_2$ and ZnS (see Figure 8). It is found that current matching at 15.3 mA/cm$^2$ is attainable under the AM1.5G solar spectrum. Note that this $J_{sc}$ is slightly higher than predicted by our model in Section 3, which indicates that In compositions below 7% in practical 1000 nm thick GaInAs subcell can be enough to attain the same $J_{sc}$ as in the standard 3JSC. A thinned GaInP the top cell could also serve to achieve current matching, but with the voltage penalty commented before.

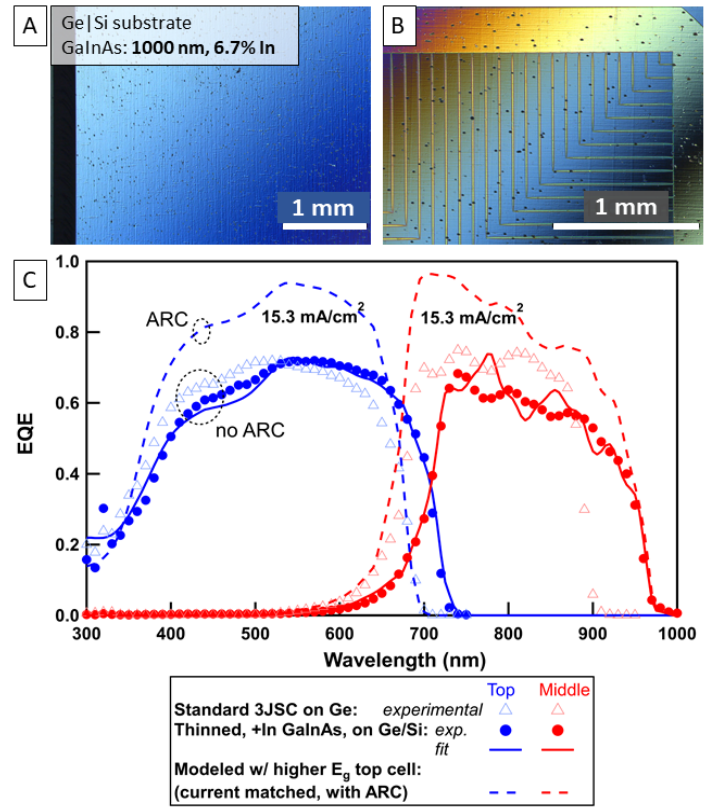

Figure 8. A-B: Nomarski microscope images of the 3JSC on Ge|Si right after growth (A) and after processing of the solar cell devices (B). C: EQE of the top and middle cells of the 3JSC implemented. The symbols correspond to the experimental results for the standard 3JSC on Ge substrates (empty triangles) and the 3JSC with thinned and 6.7% In content GaInAs middle cell on Ge|Si virtual substrate (filled circles). The solid line is a fit to the latter. The dashed line corresponds to a projection of the EQE obtained by using the fitted data but increasing the bandgap of the top cell and adding an optimized MgF$_2$/ZnS ARC.

Light J-V curves of the 3JSC are shown in Figure 9, compared to the case of standard Ge substrate. Firstly, these J-V curves demonstrate the functionality of the 3JSC developed, with a $J_{sc}$ similar to the standard design. Note that the full $J_{sc}$ increase expected when raising the In content in the GaInAs subcell is limited by the decrease of the bandgap in the GaInP top cell at the same time, which does not allow reaching the $J_{sc}$ of the standard 3JSC grown on Ge. Once the top cell bandgap is raised (or it is thinned), the $J_{sc}$ of the redesigned 3JSC should match the $J_{sc}$ of the standard 3JSC. The $V_{oc}$ drop observed has different components. Firstly, a $V_{oc}$ drop of ~185 mV is caused by the lower bandgap in the GaInAs and GaInP subcell, as estimated from EL peak measurements. Then, the effect of the TDD on the $V_{oc}$ of these subcells can be estimated to be around 350 mV [10]. An additional $V_{oc}$ drop can be attributed to the lower quality in the 6.7% In material of the GaInAs subcell discussed before (estimated 78 mV drop), which is probably affecting the GaInP top cell for the remaining of the $V_{oc}$ difference.

These results demonstrate a crack-free as-grown prototype 3JSC on Ge|Si virtual substrate. The final solar cell devices exhibit some cracks, but with promising functionality, illustrating the potential of this approach, as the main goal of this work. At the same time, these results emphasize on the fact that attaining an increased $V_{oc}$ will require first and foremost using a lower TDD virtual substrate, and improving the epilayer material quality, which is the main goal of our current efforts.

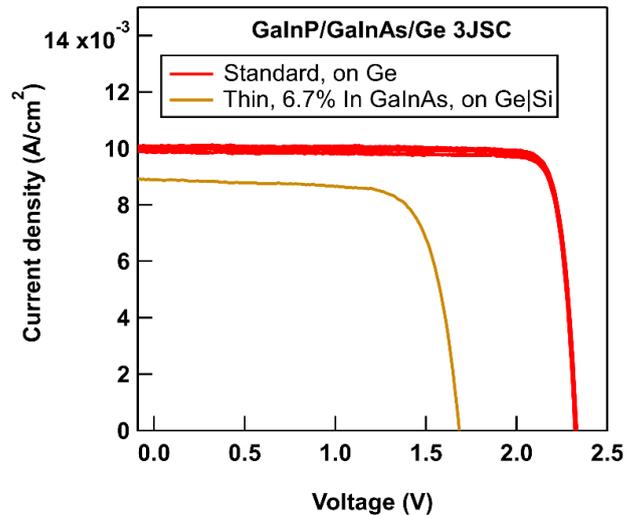

Figure 9. Light J-V measurements of the 3JSC grown on Ge|Si with thin and 6.7% In GaInAs subcell, compared to the case of a typical design on standard Ge substrate.

## 6. Summary and conclusions

The formation of cracks in GaInP/GaInAs/Ge 3JSC grown on Ge|Si virtual substrates, demonstrated to be a major challenge in our previous works, can be mitigated by thinning the structure, which we have attempted by reducing the thickness of the Ge layer in the Ge|Si template and of the GaInAs middle cell. Our theoretical calculations, based on empirical parameters obtained from standard 3JSC developed previously, indicate that a thin GaInAs middle cell with 1000 nm base layer and 8% In composition, with an (Al)GaInP top cell, should allow reproducing the performance of a standard 3JSC with a low $V_{oc}$ loss of around 3% at 1-sun and 2% for concentrations of 1000 suns. Thinning this junction is predicted to help increasing the $V_{oc}$ through a reduced recombination volume. This is demonstrated experimentally with even higher $V_{oc}$ increases than predicted theoretically in our GaInAs cells grown on Ge|Si, since they exhibit higher recombination currents than the cells grown on standard Ge substrates used for the modeling. Moreover, these thinned structures do not exhibit cracking. A higher In content up to 6.7 % in the GaInAs subcell is obtained experimentally by means of a CGB structure that replaces the usual overbuffer layer and attains a low residual strain. This thin metamorphic 6.7% In cell achieves $J_{sc}$ values similar as in cells grown lattice-matched to Ge. However, the metamorphic growth on the Ge|Si templates is found to result in an additional voltage loss, probably due to an increased TDD generated during metamorphic growth on the highly dislocated Ge|Si substrates. Full 3JSC including this GaInAs subcell are also implemented to demonstrate the concept and anticipate integration issues. Despite the total 3JSC thickness is slightly above the expected critical thickness for crack propagation, the effect of the compressive strain in the CGB used appears to mitigate the formation of cracks in the as-grown devices. In this way, we demonstrate the potential of this design concept with functional 3JSC solar cell devices that exhibit an adequate carrier collection in the top and middle cells. Once the top cell is thinned or its bandgap is raised, these solar cells should match the $J_{sc}$ of standard 3JSC grown on Ge. However, the growth on Ge|Si templates with high TDD, possibly exacerbated by the CGB, affects the quality of the top and middle cell absorber layers, giving rise to a large $V_{oc}$ loss. Further development beyond the demonstration of the concept intended for this work relies on the ability to improve the material quality and lowering the recombination currents, by using new Ge|Si templates with lower TDD and redesigning the CGB used to minimize the introduction

of additional dislocations. Once this is attained, the efficiency achievable is only about 2-3% lower than in the standard 3JSC grown on Ge substrates due to the lower voltage in the GaInAs subcell with high In content. To determine the economic worthiness of this approach, this efficiency loss will have to be assessed against the economic benefits of using Ge|Si virtual substrates, such as the lower Ge consumption and potential cost, as well as other technological advantages such as the possibility of enabling high throughput ELO processes by means of embedded porous layers in the Si substrate.

## ACKNOWLEDGEMENTS

We acknowledge technical support from J. Bautista and L. Cifuentes for device processing. Funding from AEI (project RTI2018-094291-B-I00) is gratefully acknowledged. M. Hinojosa is funded by the Spanish MECD through a FPU grant (FPU-15/03436), I. García is funded by the Spanish Programa Estatal de Promoción del Talento y su Empleabilidad through a Ramon y Cajal grant (RYC-2014- 15621) and S. Dadgostar was funded by JCYL and FEDER (Project VA283P18).